\documentclass{ifae}
\usepackage{amssymb}
\begin{document}

\begin{header}

\title{Ultra- and Extremely High Energy  Neutrino Astronomy}

\begin{Authlist}
Igor~Sokalski\Iref{ba}\Acknow{a}

\Affiliation{ba}{Istituto Nazionale di Fisica Nucleare / Sezione di Bari, 
                 Via Amendola 173, I-70126 Bari, Italy}
\Acknowfoot{a}{On leave from Institute for Nuclear Research 
               of the Russian Academy of Sciences, Moscow, Russia}
\end{Authlist}

\begin{abstract}
Scientific motivations for ultra- and extremely high energy neutrino astronomy 
are considered. Sources and expected fluxes of EHE/UHE neutrinos are briefly
discussed. Operating and planned experiments on astrophysical neutrino 
detection are reviewed focusing on deep underwater/ice Cherenkov neutrino 
telescopes.
\end{abstract} 
  
\end{header}

\section{Introduction}

From cosmic ray studies, there is a clear evidence that energies of primary 
cosmic rays extend up to enormous energies of more than $10^{20}$\,eV with 
highest energy cosmic rays detected by Fly's Eye (HiRes) collaboration
\cite{fly,hires}, Yakutsk air shower array \cite{yakutsk} and the AGASA 
experiment \cite{agasa}.
At the 
same time, the highest energy cosmic rays represent still a {\it terra 
incognita} with respect to the processes powering them. The key question of 
modern astrophysics - namely, {\it What is the nature of the cosmic high 
energy world?} - has to be considered as unsolved. There is no probe except 
for neutrino which could help us to answer this question. Electrically charged 
protons and heavier nuclei, whose arrival direction is scrambled by galactic 
and intergalactic magnetic fields, are able to point back to the sources of
their acceleration only above approximately 1--10\,EeV~\footnote{Let us remind
the energy units relevant to the discussed topic: 1\,GeV$=10^{9}$\,eV, 
1\,TeV$=10^{12}$\,eV, 1\,PeV$=10^{15}$\,eV, 1\,EeV$=10^{18}$\,eV, 
1\,ZeV$=10^{21}$\,eV, correspondingly.}. $\gamma$-rays keep the initial 
direction but the Universe is not transparent for them at energies above TeV 
range since they annihilate into electron-positron pairs in an encounter 
with a 2.7\,K cosmic microwave background photons or with infra-red radiation.
For example, $\gamma$-quantum of 1\,PeV energy can not reach us even from the 
Galaxy center (10 kpc). Neutrons are too short-live particles and they are not
in time to cross even our Galaxy before decaying if their energy is below 
several EeV. Thus, neutrino remains the only {\it i)} weak interacting; 
{\it ii)} stable; and {\it iii)} neutral probe which can reach the Earth 
(where we are able to observe it) from the cosmological distances keeping 
original direction and pointing back to the source of its origin, meeting thus
the basic requirements of {\it astronomy}.

MeV-range neutrino astronomy have been existing for forty years with two
neutrino sources identified so far, namely the Sun and Supernova SN-1987A, 
which at the moment remain the only two experimentally proved extraterrestrial
neutrino sources. Development of ultra- and extremely high 
energy~\footnote{Ultrahigh energy range (UHE) is 
$E_{\nu} =$\,30\,TeV\,--\,30\,PeV; extremely high energy range (EHE) is 
$E_{\nu} >$\,30\,PeV, respectively.} neutrino astronomy is under way, being 
still in its infancy. It started in 1960 with academician Markov's suggestion 
to use a natural basins (lakes or seas) to deploy there a large volume 
neutrino telescopes \cite{markov}. The large instrumented volume is needed 
due to the two 
basic reasons: firstly, expected fluxes of UHE/EHE neutrinos are very low and,
secondly, cross section of neutral current (NC) and charged current (CC) 
neutrino interactions 
$\nu_{l} \, N \stackrel{NC(CC) }{\longrightarrow} \nu_{l} (l) \, X$ (by which
neutrinos are supposed to be detected) is small despite its increase with the
neutrino energy. To detect neutrinos associated with highest energy cosmic 
rays one needs a kilometer scale detectors. After the first experimental steps
at the middle of the 1970th (the DUMAND project \cite{dumand}) and detection 
of the first 'underwater' atmospheric neutrino at the middle of the 1990th 
(the BAIKAL experiment \cite{1neutrino}) experimental groups and 
collaborations  moved to the next stage: 
creation of detectors with effective areas of 0.1 km$^2$ and higher with an 
ultimate goal to build neutrino telescopes with effective volumes of one
cubic kilometer. 

This talk reviews the physical goals and experimental status for ultra- and 
extremely high energy neutrino astronomy focusing, first of all, on operating 
and planned deep underwater/ice Cherenkov neutrino telescopes. 

\section{Detection Principles and Scientific Goals}

\begin{figure}[hbtp]
  \begin{center}
    \epsfig{file=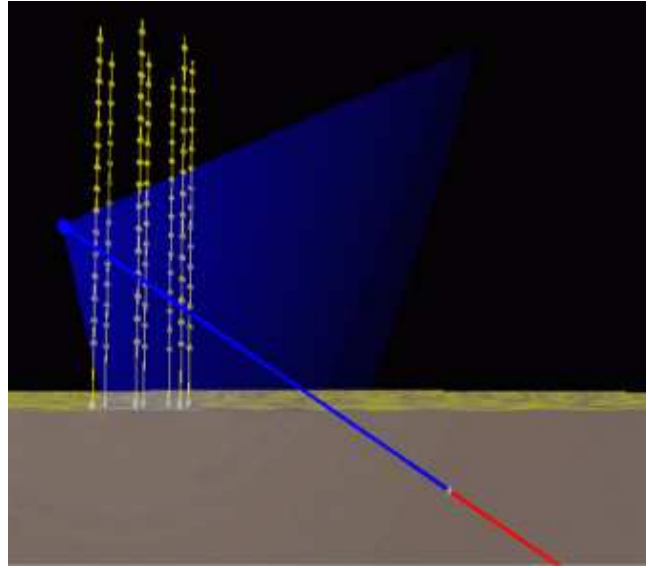,width=85mm}
    \caption{Neutrino detection in an UNT (schematic view)}
    \label{fig:principle}
  \end{center}
\end{figure}

Underwater/ice neutrino telescopes (UNTs) represent a 3-D arrays of 
photomultipliers deployed deep in the lake, ocean or in the polar ice 
at the depth of 1 to 4 kilometers to provide with a shield against the sun 
and moon light 
background and background of atmospheric muons. Detection principle is based 
on registration of the Cherenkov photons emitted by charged leptons 
(including those emitted by secondaries produced along their way in the water 
or ice and by their decay products) which are generated in CC neutrino 
interactions $\nu_{l}\,N\stackrel{CC}{\longrightarrow} l \, X$ (see 
Fig.~\ref{fig:principle}). 
Also hadronic showers produced in NC neutrino interactions 
$\nu_{l}\,N\stackrel{NC}{\longrightarrow} \nu_{l} \, X$ inside UNT sensitive 
volume can be detected by radiated Cherenkov photons. PMT hit times and 
positions provide with a possibility to reconstruct the track or shower
vertex while a charge collected on PMT anodes allows to reconstruct the
energy. 

\begin{figure}[hbtp]
  \begin{center}
    \epsfig{file=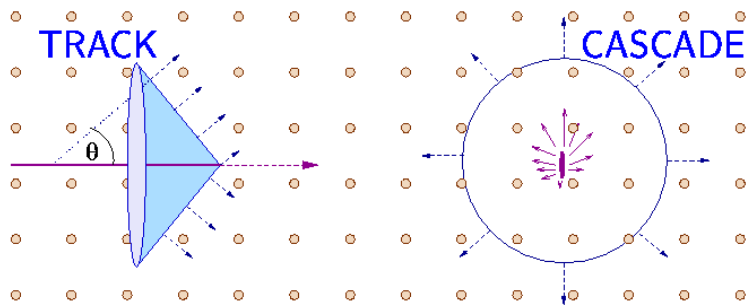,width=100mm}
    \caption{Two event topologies in an UNT (schematic view).}
    \label{fig:topo}
  \end{center}
\end{figure}
Thus, there can be two main event topologies in UNTs (Fig.~\ref{fig:topo}): 
\begin{itemize}
\item
{\bf track event} in case of muon of any energy or tau-lepton with energy 
$E_{\tau} \gtrsim$\,2\,PeV (approximately above this energy tau-lepton is able
to propagate remarkable distance before decay thanks to the Lorentz factor);
\item
{\bf shower event} in case of electron, tau-lepton with energy 
$E_{\tau} \lesssim$\,2\,PeV and NC interactions of all flavor neutrinos.
\end{itemize}
However, real events may contain both topologies. Track of UHE/EHE muon or
tau-lepton is complemented by showers produced by secondaries which are 
generated in the muon interactions: bremsstrahlung, direct $e^{+}e^{-}$-pair 
production, photonuclear interactions and knock-on electron production. With 
some probability these showers can take the major fraction of $\mu$/$\tau$ 
energy and even all the energy (first of all due to  bremsstrahlung). Thus, a 
combined topology {\bf track$\bf{+}$shower} takes place. As well as, a 
tau-lepton track with a subsequent decay create such combined topology. CC 
muon neutrino (or tau neutrino if $E_{\nu_{\tau}}$ is in multi-PeV range or 
higher) interaction within UNT sensitive volume with an hadronic shower in the
neutrino interaction vertex and subsequent charged lepton track also produces 
a combined topology which is even more complex in case of tau-lepton if it
decays inside sensitive volume providing thus with at least two showers: in 
neutrino interaction point and at the decay point (so called 'double bang'
\cite{db}). 
On the other hand, at energies below PeV range two showers at the
tau neutrino interaction vertex and at the tau-lepton decay point are so 
close 
to each other that can not be separated at reconstruction (also tau-lepton 
track can not be distinguished) and thus, such an event can be considered as
a pure shower one.   

The main goal of UHE/EHE neutrino astronomy is to determine the origin of high
energy cosmic rays. For this it is needed to detect natural flux of the 
high energy neutrinos measuring neutrino energy, directional information and
intensity. Expected sources of UHE/EHE neutrinos are as follows (more 
detailed review can be found, e.g., in \cite{halzen}):

\begin{itemize}
\item
steady sources like, e.g., Active Galactic Nuclei (AGN), Supernova Remnants 
(SNR) or microquasars; 
\item
transient sources like Gamma Ray Bursts (GBR);
\item
decay of superheavy particles or topological defects.
\end{itemize} 

Detection of UHE/EHE neutrinos and identification of their sources would 
allow to clarify the origination of UHE/EHE cosmic rays and to understand 
the processes by which the nature fills the Universe with the highest energy 
particles. 

\begin{figure}[hbtp]
  \begin{center}
    \epsfig{file=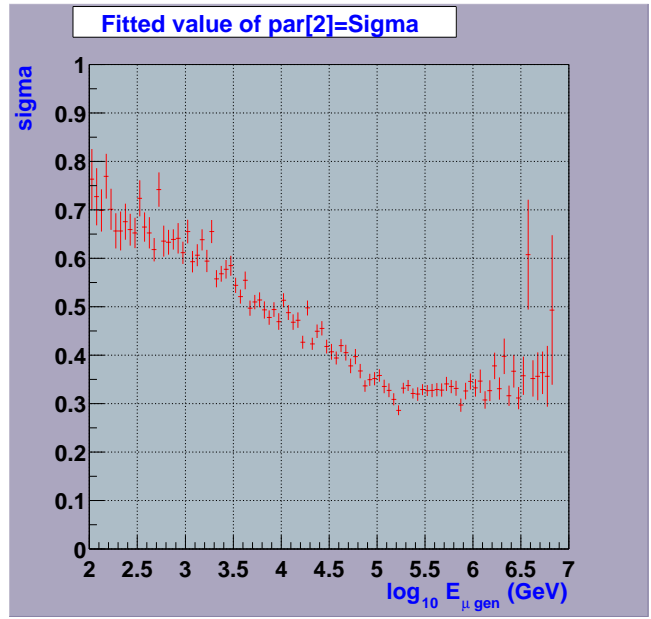,width=85mm}
    \caption{
              Energy resolution of the ANTARES 12-string detector
              (see Sec.~\ref{sec:ant})
              which is planned to be deployed by 2007
              (taken from \cite{antares1}): sigma of the distribution
              of $\log_{10}(E_{\mu}^{rec}/E_{\mu}^{t})$ versus 
              $E_{\mu}^{t}$.
             }
    \label{fig:enerres}
  \end{center}
\end{figure}
Accuracy of energy measurements in UNTs
is not too high. Energy reconstruction is based on the increase of emitted
Cherenkov light due to muon ($\tau$) catastrophic energy losses above
$\approx$1\,TeV. Also, amount of Cherenkov photons produced by both hadronic
and electromagnetic shower is more or less proportional to the shower energy.
But due to stochastic nature of energy losses and due to the fact that an
UNT represent a non-dense detector with PMTs spaced by typically 10-100\,m, 
UNTs can not be a good calorimeter: for instance, dispersion of the 
$\log_{10}(E_{\mu}^{rec}/E_{\mu}^{t})$ distribution (where $E_{\mu}^{t}$ is 
the true muon energy and $E_{\mu}^{rec}$ is the reconstructed energy, 
respectively) is around $\sigma \approx$\,0.5 at $E_{\mu} \sim$\,5\,TeV and 
$\sigma \approx$\,0.3 for $E_{\mu} \gtrsim$\,100\,TeV which means that the 
muon energy resolution is at the level of 2-3 only
(see Fig.~\ref{fig:enerres}). Besides, an additional 
un-avoided error at neutrino energy measurement comes  from the fact that 
fraction of energy that is taken by charged lepton at neutrino CC interaction
has a distribution and if neutrino interaction occurs far apart UNT sensitive
volume and, hence, shower energy can not be reconstructed, reconstructed 
charged lepton energy is not a good estimator for the neutrino energy. 

\begin{figure}[hbtp]
  \begin{center}
    \epsfig{file=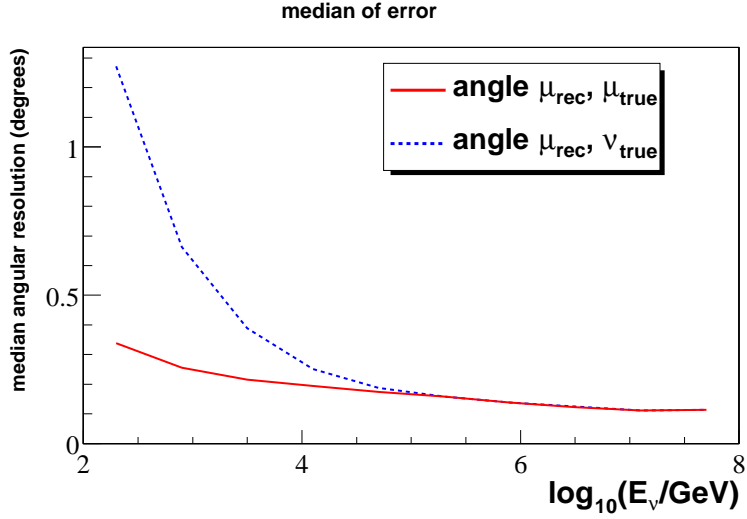,width=105mm}
    \caption{
              Angular resolution of the 12-string ANTARES detector versus
              $E_{\nu}$ (taken from \cite{antares1}): median of the 
              distribution of the angle in space between the reconstructed
              muon track and the true muon track (solid) or the parent
              neutrino track (dashed). Below $E_{\nu}\approx$\,10\,TeV 
              the reconstruction error is dominated by $\nu - \mu$ kinematics,
              at higher energies accuracy is limited only by PMT TTS and
              light scattering.
            }
    \label{fig:angres}
  \end{center}
\end{figure}
Angular resolution for track events at UHE/EHE range is, typically, at 
$\sim$0.1$^{\circ}$--1.0$^{\circ}$ level (Fig.~\ref{fig:angres}) and it is 
sufficient for search of point-like neutrino sources.

The first main background for neutrino events comes from down-going 
atmospheric muons and it is suppressed by putting UNT as deeper as possible to
provide with a shield of water or ice (each 1 km of water suppress the
atmospheric muon background by approximately one order of magnitude) and by 
selecting of up-going events as neutrino candidates. The second 
background is due to atmospheric neutrinos. The flux of astrophysical 
neutrinos is expected to behave like $E_{\nu}^{-2.0}$ whereas the 
atmospheric neutrino spectrum falls like $E_{\nu_{atm}}^{-3.7}$, yielding a 
better signal-to-background ratio at higher energies. Thus, atmospheric 
neutrino background can be suppressed by setting the off-line energy threshold
at the level $E_{thr} \sim$\,10-100\,TeV. 

Except for deep underwater or ice neutrino detection other techniques are 
also discussed and used (for more detailed review see \cite{spiering}): 

\begin{itemize}
\item
detection of coherent Cherenkov radio waves emitted by electromagnetic 
showers \cite{radio}; 
\item
acoustic pulses generated in matter heated by UHE/EHE cascades due to 
ionization energy losses \cite{acoustic}; 
\item
detection of neutrino interactions by horizontal air showers (both with 
traditional Earth-based large extensive air shower arrays \cite{au} and with 
satellite space-based  detectors \cite{owl}).  
\end{itemize} 
Such kind of experiments have a high energy thresholds (at EeV energy range) 
and are aimed to detection of highest energy neutrinos. Target masses  for
neutrino interaction are at the level of Giga-tons and higher providing with 
opportunity to detect weak neutrino fluxes.  

\section{Predicted Fluxes and Bounds}

All the models for generation of UHE/EHE particles can be divided roughly by
two main classes. 

\begin{figure}[hbtp]
  \begin{center}
    \epsfig{file=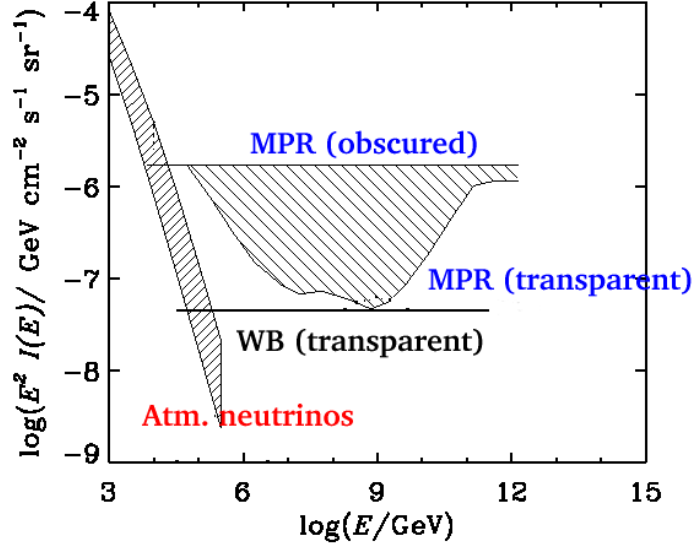,width=95mm}
    \caption{
            Waxman-Bahcall (\cite{wb}, line marked 'WB') 
            and Mannheim-Protheroe-Rachen (\cite{mpr}, lines marked 
            'MPR obscured' and 'MPR transparent') limits on diffuse
            neutrino flux. Atmospheric neutrino flux is shown at the 
            left, as well (the strip corresponds to different zenith angles). 
            }
    \label{fig:limit}
  \end{center}
\end{figure}

{\it Bottom-up} models consider initially low energy 
particles which are accelerated up to UHE/EHE, typically, by shock waves 
propagating in accretion disks around black holes or along the extended jets 
emitted perpendicularly to the disk. Neutrino are supposed to be produced in 
decays of mesons which are generated by interaction of accelerated particles 
with surrounding matter or photon fields. Such models predict $E_{\nu}^{-2.0}$
behavior of neutrino spectrum. By normalization of the neutrino flux to the 
known cosmic ray flux  one can obtain an upper bound of 
$dN/dE_{\nu}\sim 5 \times 10^{-8} E_{\nu}^{-2}$ 
GeV$^{-1}$ cm$^{-2}$ s$^{-1}$ sr$^{-1}$ (Waxman-Bahcall limit \cite{wb}) to 
the neutrino flux integrated over all possible sources or {\it diffuse} 
neutrino flux (Fig.~\ref{fig:limit}). More
detailed consideration which involves, in particular, the source transparency,
leads to bounds at the level between the Waxman-Bahcall limit and 
$dN/dE_{\nu}\sim 10^{-6} E_{\nu}^{-2}$ GeV$^{-1}$ cm$^{-2}$ s$^{-1}$ sr$^{-1}$
(\cite{mpr}, 'MPR obscured' and 'MPR transparent' in Fig.~\ref{fig:limit}).
The last flux value more or less corresponds to the best experimental
limits set by the moment on diffuse neutrino flux.

In so called {\it top-down} scenarios particles are not accelerated but, 
instead, are
born with high energies in decays of super-massive particles which generate
UHE/EHE nucleons, $\gamma$-rays and neutrinos. 

Different predictions for neutrino fluxes generated in different sources
(see, e.g., \cite{proth})
lead to expected flux at the Earth at the level of $\sim$\,100 event yr$^{-1}$ 
km$^{-2}$ above $E_{\nu}>$\,10\,TeV.

\section{Underwater/ice Neutrino Projects}

\begin{figure}[p]
  \begin{center}
    \epsfig{file=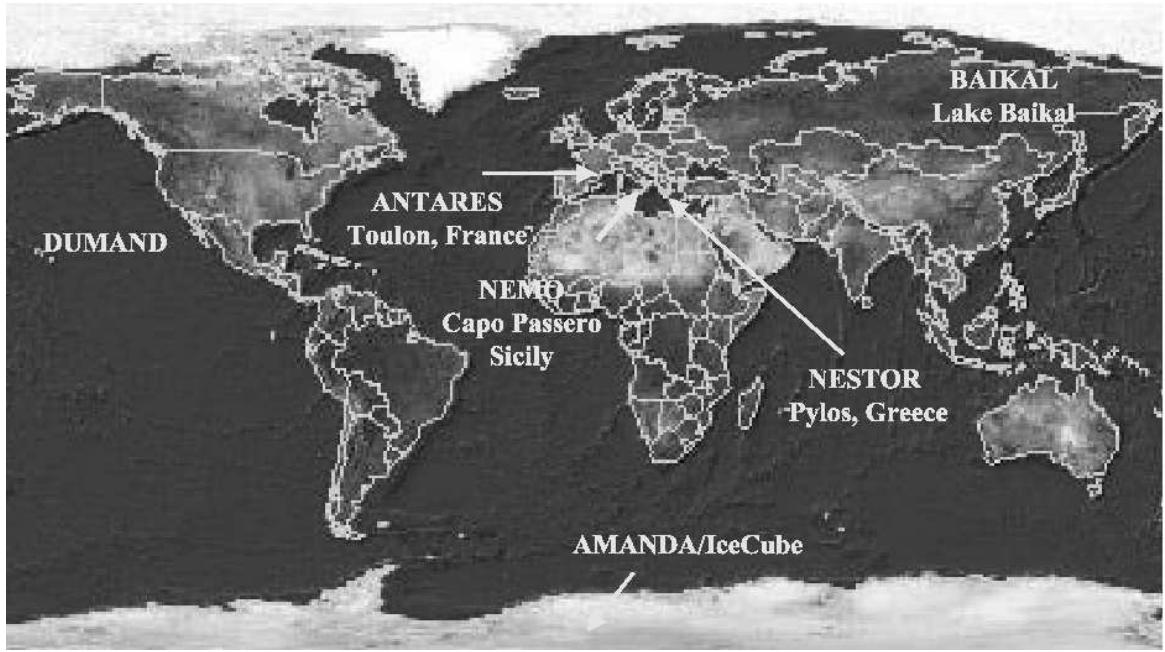,width=154mm}
    \caption{The neutrino telescope world map}
    \label{fig:map}
  \end{center}
\end{figure}

The neutrino telescope word map is shown in Fig.~\ref{fig:map}.

\begin{figure}[p]
  \begin{center}
    \epsfig{file=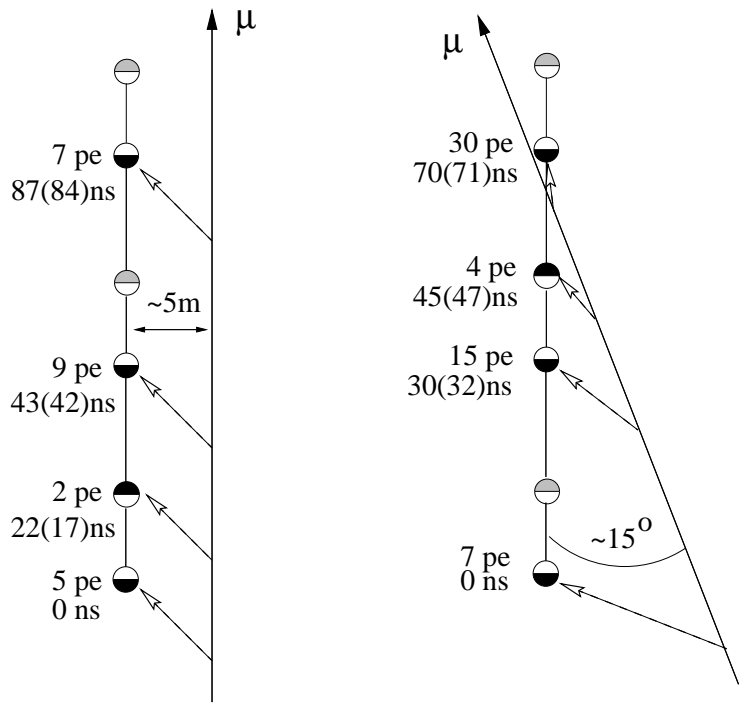,width=93mm,angle=-90}
    \caption{The first two atmospheric neutrinos detected underwater
             in the Baikal experiment in 1994. The hit PMTs are marked
             in black. Numbers give the measured amplitudes (in 
             photoelectrons) and measured (expected) times with respect 
             to the first hit channel.}
    \label{fig:twoneu}
  \end{center}
\end{figure}
\begin{figure}[hbt]
  \begin{center}
    \begin{tabular}{cc}
    \epsfig{file=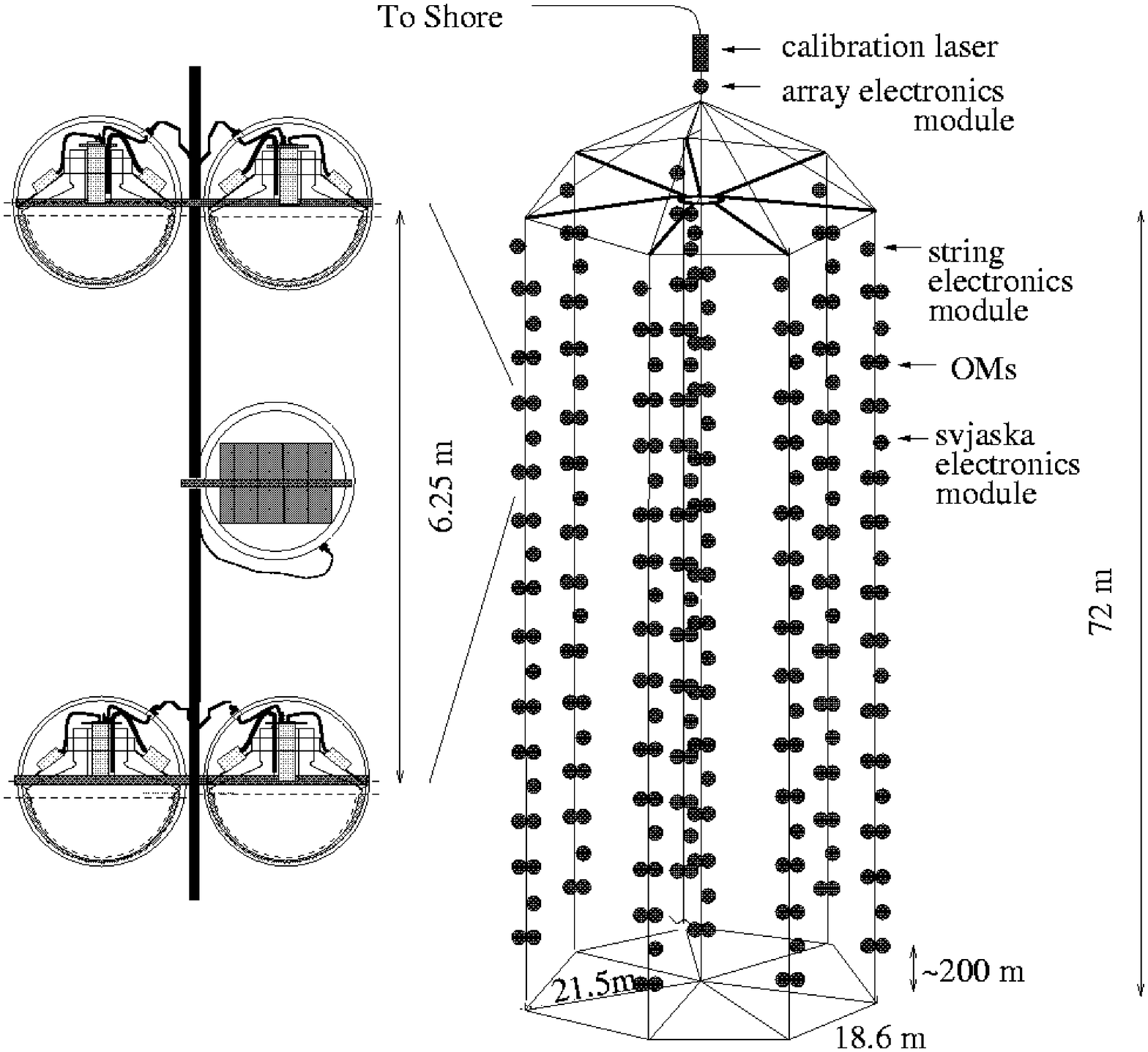,height=86mm}
    &
    \epsfig{file=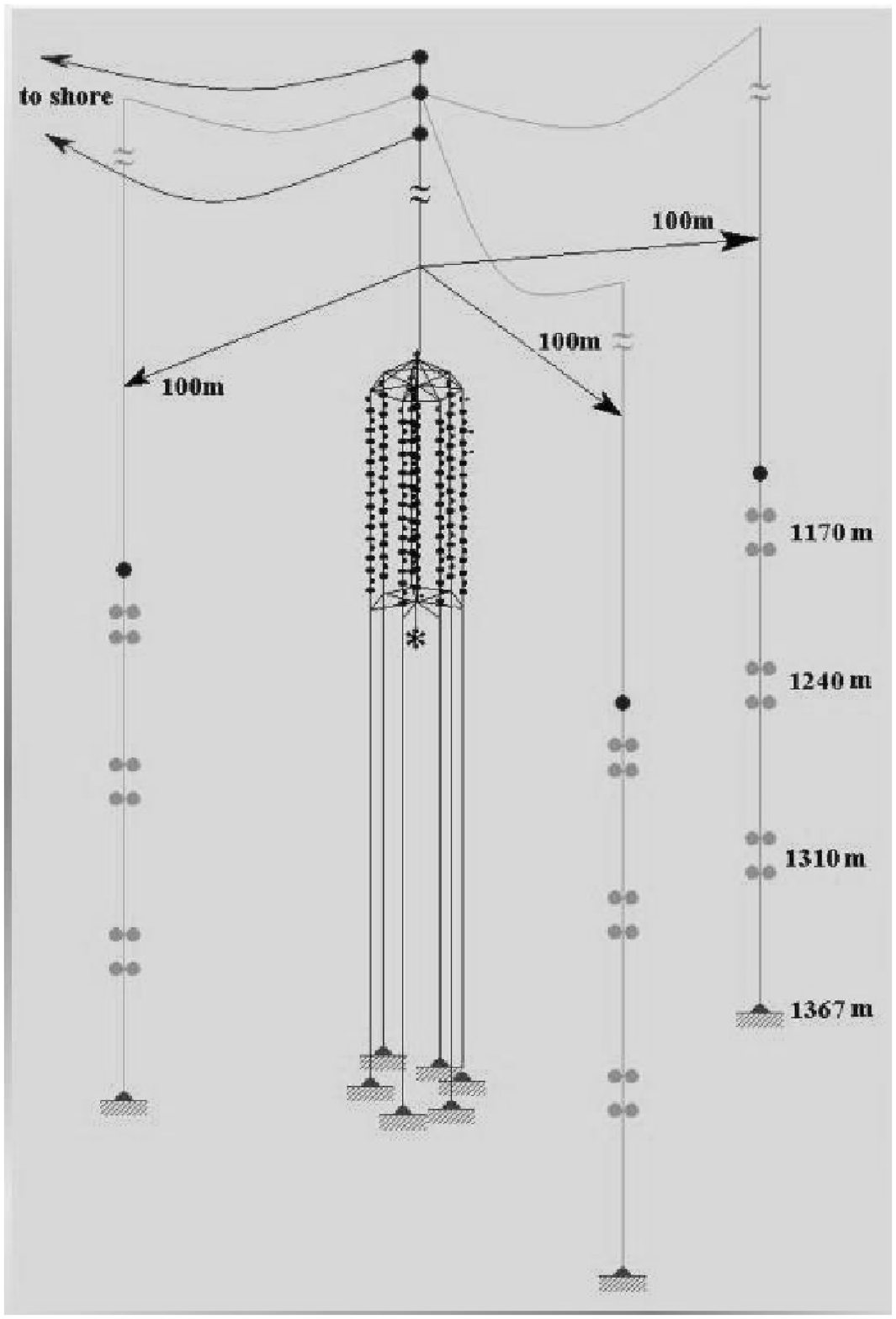,height=86mm}
    \\
    \end{tabular}
    \caption{The Baikal neutrino telescope NT-200 (left) and its planned 
             upgrade NT-200$+$ (right)}
    \label{fig:bai1}
  \end{center}
\end{figure}

\subsection{DUMAND}

The first project for deep underwater Cherenkov neutrino detection, DUMAND 
(\underline{D}eep \underline{U}nderwater \underline{M}uon and 
\underline{N}eutrino \underline{D}etector \cite{dumand}) existed from about 
1976 through 1995. The goal was the construction of the detector, to 
be placed at 4800 m depth in the Pacific Ocean off Keahole Point on the Big 
Island of Hawaii. Many preliminary studies were carried out, from technology 
to ocean optics. A prototype vertical string of instruments suspended from a 
special ship was employed to demonstrate the technology, and measure the 
cosmic ray muon flux in the deep ocean. 
The DUMAND hardware was donated to the NESTOR Project in Greece
(Sec.~\ref{sec:nes}), and may yet 
be employed there. Although the DUMAND project was canceled in 1995, it 
stimulated a lot the development of underwater technique for neutrino detection.

\subsection{Baikal}

The Baikal neutrino detector is located at a depth of 1100\,m in Siberian Lake
Baikal. The experiment started in early 80th, the first stationary 
single-string detectors equipped with 12-36 PMTs were put in operation in 
1984-86. In 1993 the Baikal collaboration was the first to deploy pioneering 
3-D underwater array consisting of 3 strings (as necessary for full 
spatial reconstruction). In 1996 the first atmospheric neutrinos detected
underwater (see Fig.~\ref{fig:twoneu}) were reported \cite{1neutrino}. 

Since 1998 8-string NT-200 detector equipped with 192 15$^{``}$ PMTs is taking
data (Fig.~\ref{fig:bai1}). An upgrade (NT-200$+$) is under construction and 
it is planned to be completed by 2005. NT-200$+$ will consist of NT-200 
surrounded by 3 additional strings placed 100\,m apart and it is optimized for
diffuse neutrino flux detection. The current limit on diffuse neutrino flux 
set by the Baikal experiment is
$dN/dE_{\nu}\sim 1.3 \times 10^{-6} E_{\nu}^{-2}$ GeV$^{-1}$ cm$^{-2}$ 
s$^{-1}$ sr$^{-1}$ for energy range 10\,TeV\,$\le E_{\nu} \le$\,10\,PeV
(assuming $E_{\nu}^{-2}$ neutrino spectra).
Besides, limits on magnetic monopole flux, Q-ball flux, 
results on search of neutralinos in the core of the Earth,  measurements on 
atmospheric muons and neutrinos were reported \cite{baikalres}.

\subsection{AMANDA/IceCube}

The first antarctic detector AMANDA-B10 was put into operation at the beginning
of 1997. It consists of 302 PMTs deployed at a depth 1500-2000\,m. AMANDA 
(\underline{A}ntarctic \underline{M}uon and \underline{N}eutrino 
\underline{D}etector \underline{A}rray) collaboration uses 3\,km thick ice 
layer at the geographical South Pole. Holes are drilled with hot water and 
then strings with PMTs are frozen into the ice. 
\begin{figure}[hbtp]
  \begin{center}
    \epsfig{file=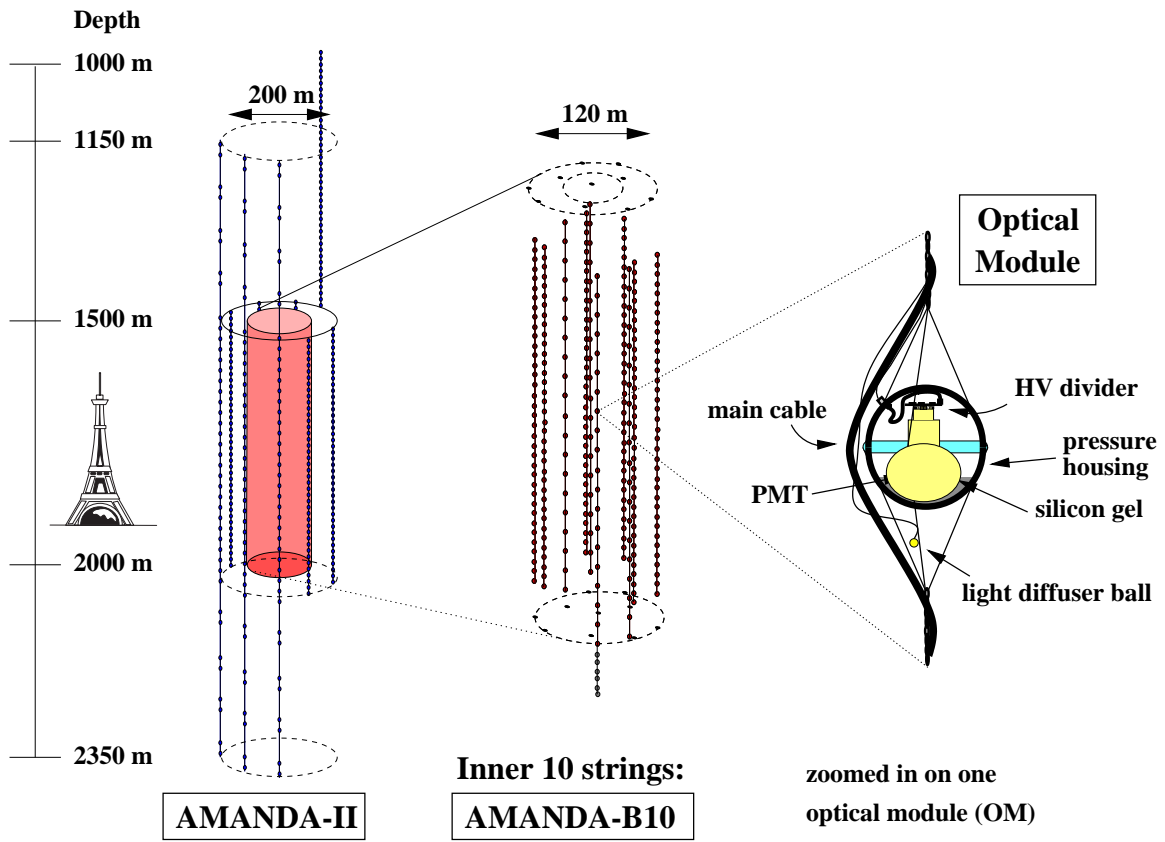,width=153mm}
    \caption{The AMANDA detector. The scheme is illustrated by Eiffel
             tower at the left. Each dot represents an optical module
             with PMT.}
    \label{fig:amanda}
  \end{center}
\end{figure}
In January 2000 deployment
of additional 9 strings was completed and since that time AMANDA-II detector 
is in operation with 677 PMTs at 19 strings (see Fig.~\ref{fig:amanda}). The 
unique feature of AMANDA is that it continuously works in coincidence with 
surface air shower experiment SPASE \cite{space} which allow to calibrate the 
angular resolution.
Given a $E_{\nu}^{-2}$ benchmark neutrino spectral shape, limits 
$dN/dE_{\nu}\sim 1.5 \times 10^{-6} E_{\nu}^{-2}$ GeV$^{-1}$ cm$^{-2}$ 
s$^{-1}$ sr$^{-1}$ and
$dN/dE_{\nu}\sim 0.86 \times 10^{-6} E_{\nu}^{-2}$ GeV$^{-1}$ cm$^{-2}$ 
s$^{-1}$ sr$^{-1}$ are set on diffuse neutrino flux in the ranges 
1\,PeV\,$\le E_{\nu} \le$\,3\,EeV and 50\,TeV\,$\le E_{\nu} \le$\,5\,PeV,
respectively \cite{amdif}. 
Estimated sensitivity to the point-like neutrino sources is
at the level of expected neutrino fluxes from AGNs Mrk421 and Mrk501
\cite{stecker}, as well as from microquasar SS433 for a specific model 
\cite{distefano}. Also results on atmospheric muons and neutrinos, WIMP
and magnetic monopoles search, search for supernovae bursts, primary
cosmic ray composition have been published by the AMANDA collaboration 
\cite{amanda}.

\begin{figure}[hbtp]
  \begin{center}
    \epsfig{file=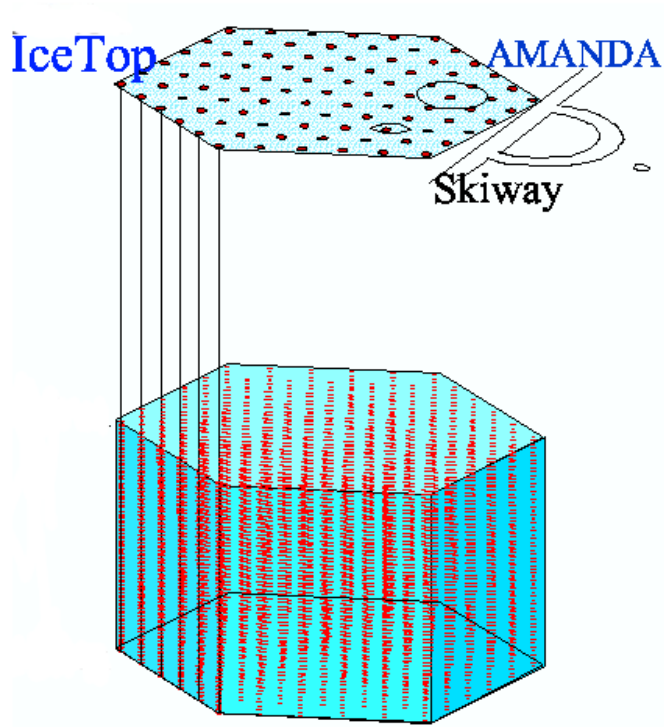,width=96mm}
    \caption{Schematic view on the future IceCube detector}
    \label{fig:icecube}
  \end{center}
\end{figure}
As a next step of development of the neutrino observatory at the South Pole
creation of neutrino telescope with instrumented volume of 1\,km$^{3}$ 
(IceCube) is foreseen \cite{icec}. It will consist of 4800 PMTs deployed on 80
vertical strings (each of 60 PMTs) at the depth from 1400\,m  to 2400\,m. 
The distance between strings is 125\,m, the distance between PMTs along the 
strings 16\,m. Existing detector AMANDA-II will be integrated to IceCube. 
Fig.~\ref{fig:icecube} gives a schematic view of IceCube and its position
with respect to AMANDA-II and the air shower array. Deployment operations at
the South Pole will begin in late 2004 and detector will be completed by
2010. With 45,000 atmospheric neutrinos recorded per year the ultimate
sensitivity to an extraterrestrial $E_{\nu}^{-2}$  neutrino flux after 
3 years of data taking is 
$dN/dE_{\nu}\sim 3(10) \times 10^{-9} E_{\nu}^{-2}$ GeV$^{-1}$ cm$^{-2}$ 
s$^{-1}$ sr$^{-1}$ (the first number refers to the 90\% limit and
the second one to the 5$\sigma$ sensitivity). This is lower compared to 
Waxman-Bahcal and Mannheim-Protheroe-Rachen upper bounds \cite{wb,mpr} and 
to the most popular predictions on diffuse neutrino flux which are based on 
different models \cite{proth}.

\subsection{NESTOR}
\label{sec:nes}

\begin{figure}[hbtp]
  \begin{center}
    \epsfig{file=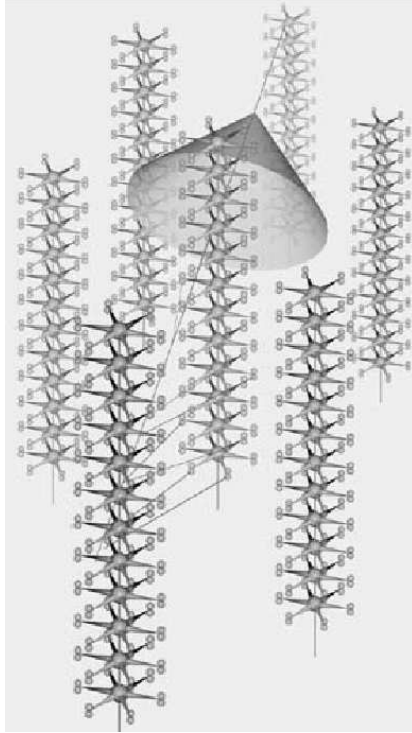,width=55mm}
    \caption{Schematic view of the NESTOR detector}
    \label{fig:nestord}
  \end{center}
\end{figure}

NESTOR (\underline{N}eutrino \underline{E}xtended \underline{S}ubmarine 
\underline{T}elescope with \underline{O}ceanographic \underline{R}esearch)
\cite{nestor}
will be deployed in the Mediterranean Sea, near Pylos (Greece) at 4\,km depth.
It is planned to be 'tower based detector' (Fig.~\ref{fig:nestord}). 
Each tower consists of 12 hexagonal floors spaced by 30\,m with 6 pairs of
up-down looking 15$^{``}$ PMTs each. The diameter of the floor is 32\,m.
The effective area of the tower with respect to TeV-range muons is about
0.02\,km$^{2}$. The NESTOR collaboration has passed through a long phase
of site evaluation and technology tests. 
An 28\,km electro-optical cable was put on the seafloor
to connect the detector and shore station in 2000 and it was repaired in 2002.
In March, 2003 a 'prototype floor' equipped with 12 PMTs was deployed.
Over 5 millions of muon triggers were recorded during its operation. 

\subsection{ANTARES}
\label{sec:ant}

The ANTARES project \cite{antares} 
(\underline{A}stronomy with a \underline{N}eutrino 
\underline{T}elescope and \underline{A}byss environmental 
\underline{RES}earch) was formed in 1996. 
In 1996-99 an intense R\&D program was performed. The deployment and recovery 
technologies, electronics and mechanical structures were developed and tested
with more than 30 deployments of autonomous strings. The environmental 
properties at the detector site were investigated. ANTARES R\&D program 
culminated with deployment and 8 month operation of a 350\,m length 
'demonstrator string' (November 1999 - July 2000) instrumented with 7 PMTs at 
a depth of 1100\,m, 40 km off the coast of Marseille. The string was 
controlled and read out via 37\,km-long electro-optical cable connected to the
shore station. $\sim$5$\cdot$10$^{4}$ seven-fold coincidences from atmospheric
muons were recorded. The angular distribution of atmospheric muons was 
reproduced and the fraction of multi-muon events was found to be 
in agreement with expectation. 

After extensive R\&D program the 
collaboration moved into construction of a 12-string detector in the 
Mediterranean Sea at 2400 m depth, $\sim$40 km off-shore of La Seyne sur Mer, 
near Toulon (42$^{\circ}$50$^{'}$\,N, 6$^{\circ}$10$^{'}$\,E).
\begin{figure}[hbtp]
  \begin{center}
    \epsfig{file=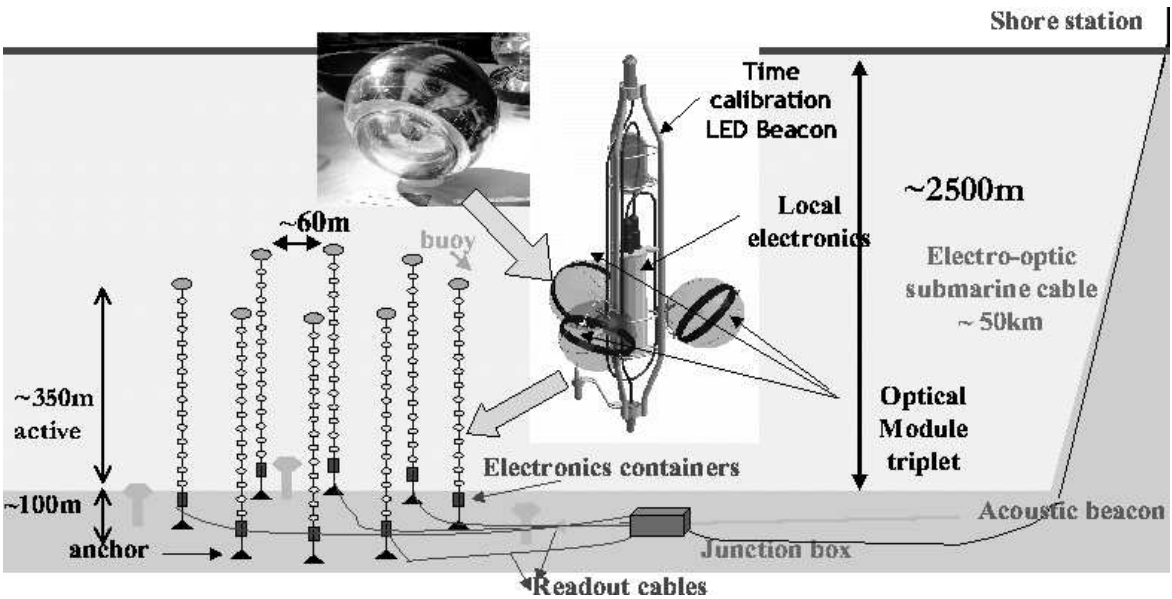,width=155mm}
    \caption{Schematic view of the ANTARES 12-string detector}
    \label{fig:antaresd}
  \end{center}
\end{figure}
Each string will be  instrumented with 75 PMTs housed in glass
spheres (see Fig.~\ref{fig:antaresd}).
PMTs are grouped in triplets at 25 levels separated by 14.5\,m. 3 PMTs
in each triplet are oriented at 45$^{\circ}$ to the nadir. Strings are 
separated from each other by $\sim$70\,m. All the strings are connected to a 
Junction Box (JB) by means of electro-optical link cables. The JB is connected
to the shore station by a 50\,km length 48-fiber electro-optical cable. 
Undersea connections are performed with a manned submarine. 
The deployment of the detector is planned for 2005-2007.
The important milestones that have been achieved by the collaboration are: 

\begin{itemize}
\item the electro-optical cable connecting detector and shore 
station was deployed in October 2001; 
\item the industrial production of 900 OMs started
in April 2002; 
\item since December 2002 the JB is in communication with the
shore station; 
\item  in December 2002 and February 2003 the 'prototype 
instrumentation string' and the 'prototype detection string' (equipped with 15
OMs) were successfully deployed (recovered in May and July, 2003, 
respectively); 
\item in March 2003 both strings were connected to JB with 
the Nautile manned submarine and data taking started. 
\end{itemize}

The sensitivity of the 
detector to diffuse neutrino fluxes achieved by rejecting the background with 
an energy cut of $E_{cut}=$\,50\,GeV  allows to  reach Waxman-Bahcall 
limit in 3 years. The ANTARES sensitivity for point-like source searches 
(90\% C.L.) assuming $E^{-2}$ differential $\nu$ flux is in the range 
4$\div$50$\cdot$10$^{-16}$\,cm$^{-2}$\,s$^{-1}$ (depending on the source 
declination) after 1 year, which gives a real hope to detect a signal from 
the most promising sources.

The deployment of the 
ANTARES neutrino telescope can be considered as a step toward 
the creation of a 1 km$^3$ detector in the Mediterranean Sea.

\subsection{NEMO}

NEMO \cite{nemo} (\underline{NE}utrino sub\underline{M}arine 
\underline{O}bservatory) is an R\&D project of the Italian National
Institute for Nuclear Physics (INFN) for 1\,km$^{3}$ neutrino underwater 
telescope to be deployed in Mediterranean Sea near Capo Passero, Sicily,
at the depth of 3500\,m where transparency  and other water parameters are 
optimal. At the first stage (1998-2000) the NEMO collaboration performed
an intensive search program (more than 20 sea campaigns) to determine the
optimal site for the future detector. Also R\&D program on materials, PMTs
and mechanical components of the detector were performed. At the second stage 
which started in 2002, the advanced R\&D and prototyping is done. The 
laboratory which is connected with test site of-shore Catania by 28\,km
electro-optical cable is used for this purpose. The overall number of 
PMTs in NEMO detector may lay between 7000 and 10000.

\section{Conclusions}

{\it
\begin{center}
\rule{75mm}{0mm} ``...because then we might find something\\[0mm] 
\rule{83mm}{0mm}   that we weren't looking for, which might\\[0mm] 
\rule{85mm}{0mm} be just what we were looking for, really...''\\[0mm] 
\end{center}
}

\vspace{1mm}
\rule{105mm}{0mm} A.~Milne, ``Winnie-The-Pooh''~\cite{pooh}

\begin{figure}[hbtp]
  \begin{center}
    \epsfig{file=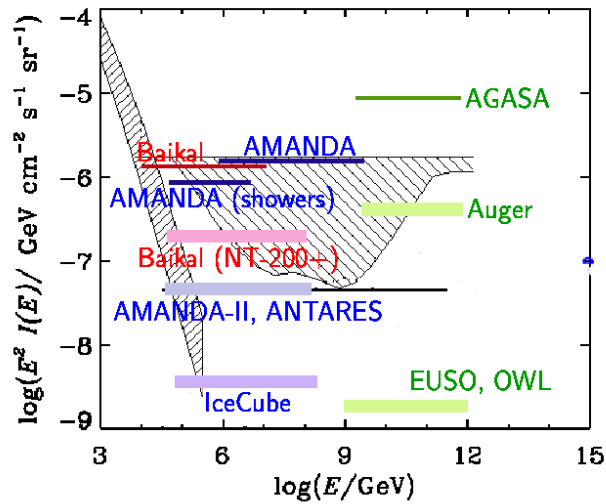,width=86mm}
    \caption{
             Present limits (horizontal solid lines) on diffuse 
             extraterrestrial neutrino flux and expected sensitivity
             to neutrino fluxes (horizontal strips), assuming 
             $E^{-2}$ behavior of neutrino spectrum, for operating
             and planned UHE/EHE neutrino experiments. Upper bounds
             on diffuse neutrino fluxes and atmospheric neutrino spectrum 
             are given, as well (see caption to Fig.~\ref{fig:limit}).
            }
    \label{fig:limit2}
  \end{center}
\end{figure}

Thus, during the next decade several 0.1--1.0 kilometer scale underwater 
(ice) Cherenkov neutrino telescopes will be put into operation both in 
Southern and North Earth hemispheres being complemented by other technique 
detectors (radio, acoustic, air showers). 
Expected sensitivity of these detectors to extraterrestrial neutrino fluxes 
(see Fig.~\ref{fig:limit2}) gives a hope to open a new era in UHE/EHE 
neutrino astronomy by detection of high energy neutrino signal.
Both discovering and {\it not} 
discovering of extraterrestrial UHE/EHE neutrinos will help to solve
one of the oldest astrophysical puzzle - {\it origin of highest energy
cosmic rays}. But hopefully, it will also lead
to discovery of new unexpected phenomena and setting new puzzles 
that will have to be solved with next generation detectors and next 
generation of scientists.

\end{document}